\documentclass[aps,prb,twocolumn,showpacs,showkeys]{revtex4}
\usepackage{amsfonts}
\usepackage{amsmath}
\usepackage{amssymb}
\usepackage{graphicx}

\begin{document}

\title{Theory of semiconductor quantum-wire based 
single- and two-qubit gates}

\author{Tobias Zibold}
\affiliation{Walter Schottky Institute, Technische 
Universit\"{a}t M\"{u}nchen, 85748
Garching, Germany}

\author{Peter Vogl}
\affiliation{Walter Schottky Institute, Technische 
Universit\"{a}t M\"{u}nchen, 85748
Garching, Germany}

\author{Andrea Bertoni}
\affiliation{CNR-INFM National Research Center on NanoStructures 
and BioSystems at Surfaces (S3), 41100 Modena, Italy}



\begin{abstract}
A GaAs/AlGaAs based two-qubit quantum device that allows the
controlled generation and straightforward detection of entanglement by
measuring a stationary current-voltage characteristic is proposed. We
have developed a two-particle Green's function method of open systems
and calculate the properties of three-dimensional interacting
entangled systems non-perturbatively.  We present concrete device
designs and detailed, charge self-consistent predictions. One of the
qubits is an all-electric Mach-Zehnder interferometer that consists of
two electrostatically defined quantum wires with coupling windows,
whereas the second qubit is an electrostatically defined double
quantum dot located in a second two-dimensional electron gas beneath
the quantum wires. We find that the entanglement of the device can be
controlled externally by tuning the tunneling coupling between the two
quantum dots.

\end{abstract}
\startpage{1}
\maketitle

\section{Introduction}  \label{intro}

Semiconductor based concepts for quantum information processing
promise a high degree of scalability.  However, the implementation of
even a single qubit proves to be more difficult in semiconductors than
in alternative approaches.~\cite{CY1995, DRBH1995, CZ1995, DiV1995,
GC1997, BCJD1999, NielsenChuang} The reasons lie in the short
decoherence times and strong interactions between elementary
excitations in solids.~\cite{KSSCEBAFZ2005} While a 7-qubit NMR
quantum computer was realized already in 2001,~\cite{VSBYSC2001} the
first two-qubit quantum operation in semiconductors has been
demonstrated with spin qubits only
recently.~\cite{LDiV1998,KBTVNMKV2006,PJTLYLMHG2005}

At liquid helium temperatures and below, electrons in a GaAs/AlGaAs
two-dimensional electron gas (2DEG) show mean free paths as well as
phase-relaxation lengths of the order of $10-20~\mathrm{\mu
m}$ which are remarkably long distances.~\cite{YSUH1991} Indeed, 
Bertoni \textit{et al.}\cite{BBBJR2000} and 
Ionicioiu \textit{et al.}\cite{IAU2001} have
proposed a scheme for quantum computation in semiconductors that
exploits these long coherence lengths for electrons propagating
through quantum wires (QWRs).  In this scheme, a single electron
propagates through two parallel QWRs, that represent the qubit states
$\left\vert 0\right\rangle$ and $\left\vert 1\right\rangle$,
respectively. 
A single-qubit rotation gate can be realized by an electronic
directional coupler\cite{delAE1990,PPBFRP2005} that may consist of a
small window in the barrier between the two QWRs and is able to
transfer the wave packet from one channel to the other.  Single-qubit
structures have been studied theoretically by several
authors\cite{PA2002,HAF2001,ARSS2004,MBRR2004} who showed how the
dimensions of the coupling window can be utilized to tailor the
transfer process.  This approach allows one, at least in principle, to
apply multiple quantum gate operations to a qubit without exceeding
the relevant coherence lengths.

Universal quantum computation requires not only one-qubit but
also two-qubit gates~\cite{DiV1995} that are responsible for the 
creation of entanglement between the qubits. 
To this end, two pairs of QWRs may be brought close to one another so
that the electrons in the QWRs can interact with one another
in a controlled way via
their Coulomb repulsion.  This concept has been analyzed both for wave
packets\cite{IZR2001,BBBJR2002,BIZRJ2002,MRBR2003} as well as for
stationary states\cite{ARSS2004,RS2005,SR2004}.
The latter investigations used simple models that provide important
proofs-of-principle, but do not provide quantitative predictions of
realizable device structures that exhibit quantum gate operations.

While two-qubit devices have only been addressed theoretically in the
literature, one-qubit systems have also stimulated a number of
experiments in both lateral\cite{RBR2006a,RBR2006b} as well as
vertical\cite{FAKSA2006,SF2007} nanostructures. However, 
quantum entanglement can clearly not be exposed
in single-qubit systems as entanglement expresses the
non-separability of a multi-qubit state. Therefore, the
following step towards the realization of a prototype quantum gate
based on coupled QWRs consists of a two-qubit device.  Also, we should
keep in mind that conceivable two-qubit experiments need 
realistic estimations of an
entanglement witness that is experimentally accessible and 
a clear signature of the correct coherent functioning of the
directional coupler.

The aim of the present work is to theoretically study realistic
single- and two-qubit quantum gates, by proposing conceptually simple
and experimentally realizable novel semiconductor devices for quantum
information processing.  The system proposed is based on ballistic
GaAs/AlGaAs QWRs and allows for the controlled generation and
detection of entanglement between an all-electric Mach-Zehnder
interferometer and an electrostatically defined single-electron double
quantum dot.  The Mach-Zehnder interferometer is realized by two
electrostatically defined QWRs that are connected by two coupling
windows. We model the electrons that are propagating through the 
interferometer by stationary scattering states, and our qubit state 
will be defined accordingly. In fact, the original proposal of Bertoni
\textit{et al.} involves the injection of electron wave packets into
the pair of QWRs. However, we chose here a time-independent approach 
since, (i) it is
equivalent to the time-dependent one when the spatial dimension of the
wave packet tends to be larger than the device (indeed, this is the
case in our devices due to the small source/drain biases leading to a
well defined kinetic energy of the electron), (ii) the structures that
implement the quantum gates operate in the same way, (iii) it allows
us to compute realistic estimations of the charge-self-consistent
ballistic $I$-$V$ characteristics of the device. 

In our calculations, we include the detailed charge-self consistent
three-dimensional device geometry, material composition, doping
profile and bias voltage. Importantly, we demonstrate that the 
all-electric Mach-Zehnder interferometer can function as 
a fully controllable single-qubit gate for experimentally 
attainable parameters. 
In addition, we have developed a Green's function method for the
quantitative analysis of the entangled Mach-Zehnder and double quantum
dot system that includes the Coulomb interaction between the two
qubits non-perturbatively. In order to gain better qualitative insight
into the numerical results, we interpret them in terms of an
analytical model that reproduces the computational results
qualitatively.  Taking both results together, we are able to 
show that the degree of entanglement can be
related to the DC $I$-$V$ characteristics of the interferometer and
that the Mach-Zehnder double quantum dot device can be employed as an
externally controllable two-qubit gate.

This paper is organized as follows.  In Sec.~\ref{method}, we
introduce a novel method for the quantum mechanical calculation of
the ballistic current through an interacting two-particle system.
This includes the calculation of the electronic structure and the
determination of the ballistic transmission characteristics of the
realistic three-dimensional nanostructure. In 
Sec.~\ref{numericaldetails}, we discuss numerical details of the
method. In Sec.~\ref{analyticalmodel} we introduce a simplified 
analytical scheme able to gather the essential features of the 
two-qubit device. The results obtained with this approach 
will be compared, in
the following section, with the calculations of the realistic device.
Sec.~\ref{results} focuses on the results and the discussion of the
ballistic current through the proposed GaAs/AlGaAs single- and
two-qubit QWR devices. In the same section, the degree of entanglement
between the two qubits is evaluated.
The paper concludes with final remarks and a summary in
Sec.~\ref{conclusions}.

\section{Method}  \label{method}

In this section, we present a novel method for the calculation of the 
stationary ballistic current of two interacting particles in 
realistic three-dimensional nanostructures. We restrict 
ourselves to the case where only one of the two particles 
(say, particle 1) contributes to the current while the second particle 
(particle 2) is assumed to be bound. Importantly, this 
method accounts for many-particle effects non-perturbatively which is an 
important prerequisite of studying entanglement and, consequently, 
predicting solid-state based quantum information processing. The 
presently developed scheme generalizes the contact block 
reduction (CBR) method for single-particle ballistic currents that 
we have developed previously.~\cite{MSV2003, MVSZV2005} 
In analogy to the CBR method, we proceed in two steps that we will
describe in this section.  First, we determine the equilibrium
electronic structure of a closed two-particle system, obtained by
substituting the leads of our device with closed boundary conditions.
Then, in the second step, we calculate the current carrying states of
the open two-particle system.  In fact, the closed calculation only
provides a convenient basis for the computation of the scattering
states.  For the sake of efficiency and without loss of generality, we
perform the former calculation in the Hartree approximation, by
solving two coupled Schr\"{o}dinger and Poisson equations.  Then, we
use single-particle product states as the basis for the scattering
states of the fully interacting two-particle device.

\subsection{Electronic structure of particle 1}

We consider a single-particle, single-band, effective mass 
Schr\"{o}dinger equation for this electron, and include the electrostatic 
Hartree potential in the absence of the second particle. The closed 
system is represented by a Hermitian Hamiltonian 
with von Neumann boundary conditions for the wave functions at the 
contacts (vanishing normal derivative).~\cite{MSV2003, MVSZV2005} The 
electronic structure of the closed device is calculated 
self-consistently. To this end, we iteratively solve the Schr\"{o}dinger 
equation
\begin{align}
H_{1}\langle\mathbf{x}|\alpha_{i}\rangle 
& = \left[  -\frac{\hbar^{2}}
{2m}\nabla\frac{1}{m^{\ast}\left(  \mathbf{x}\right)  }\nabla+E_{\mathrm{c}
}\left(  \mathbf{x}\right)  -e\phi\left(  \mathbf{x}\right)  \right]
\langle\mathbf{x}|\alpha_{i}\rangle \notag \\
 & =E_{i}^{0}\langle\mathbf{x}|\alpha_{i}\rangle\, ,\label{SchroedingerEquation}
\end{align}
and the non-linear Poisson equation
\begin{equation}
\nabla\varepsilon_{\mathrm{r}}\left(  \mathbf{x}\right)  \varepsilon_{0}
\nabla\phi\left(  \mathbf{x}\right)  =-e\rho\left[  \phi\right]\, ,
\label{NonlinearPoissonEquation}
\end{equation}
until the electrostatic potential $\phi$ and the total charge 
density $\rho$ have reached convergence. The effective mass $m^{\ast}$, 
the conduction band offset $E_{\mathrm{c}}$, and the relative dielectric 
constant $\varepsilon_{\mathrm{r}}$ are position-dependent material 
parameters in a general
three-dimensional nanostructure that may be composed of several different
materials. The total charge density is composed of the electron density
$n\left(  \mathbf{x}\right)$ and the ionized donor density 
$N_{\mathrm{D}}^{+}\left(\mathbf{x}\right) \;,$
\begin{equation}
\rho\left(  \mathbf{x}\right)  =-n\left(  \mathbf{x}\right)  
+N_{\mathrm{D}}^{+}\left(  \mathbf{x}\right)  .
\end{equation}
The latter results from the donor density $N_{\mathrm{D}}\left( 
\mathbf{x}\right)  $, the degeneracy $g_{\mathrm{D}}$, and the donor 
energy $E_{\mathrm{D}}$ according to the Thomas-Fermi approximation,
\begin{equation}
N_{\mathrm{D}}^{+}\left(  \mathbf{x}\right)  =\frac{N_{\mathrm{D}}\left(
\mathbf{x}\right)  }{1+g_{\mathrm{D}}
e^{\left(  E_{\mathrm{F}}-E_{\mathrm{D}}\right)  /k_{\mathrm{B}}T}} \;.
\end{equation}
The electron density is calculated quantum mechanically by occupying the 
electronic states according to the Fermi-Dirac statistics
\begin{equation}
n\left(  \mathbf{x}\right)  =\sum_{i}\left\vert \langle\mathbf{x}|\alpha
_{i}\rangle\right\vert ^{2}f\left(  \frac{E_{i}^{0}-E_{\mathrm{F}}
}{k_{\mathrm{B}}T}\right)  .
\end{equation}
Here, $T$ denotes the temperature, $k_{\mathrm{B}}$ is Boltzmann's 
constant, $E_{\mathrm{F}}$ denotes the Fermi level, and $f$ is the Fermi 
distribution function.

\subsection{Electronic structure of particle 2}

For the second particle, we will focus on an electron in a closed 
double quantum dot that can tunnel between these dots. The two lowest 
lying quantum states can be described by an effective Hamiltonian, that 
takes into account the energy splitting $\Delta$ between the ground 
states of the isolated quantum dots and the tunneling coupling $t$ 
between the two quantum dots. This two-by-two Hamiltonian is given by
\begin{equation}
H_{2}|Y\rangle=
\begin{pmatrix}
-\frac{\Delta}{2} & -\frac{t}{2}\\
-\frac{t}{2} & \frac{\Delta}{2}
\end{pmatrix}
|Y\rangle=E_{Y}|Y\rangle. \label{ModelHamiltonianQuantumDots}
\end{equation}
Typical values for $\Delta$ and $t$ for lateral semiconductor quantum 
dots are of the order of $10~\mathrm{\mu eV}$.~\cite{WFEFTK2003} We 
use this model Hamiltonian for the subsystem of the second particle 
and represent the charge distribution of the electron in the ground 
state of each (isolated) quantum dot 
by a point charge centered at $\mathbf{x}_{0}$ and $\mathbf{x}_{1}$, 
respectively. The use of a model Hamiltonian rather than a realistic 
three-dimensional device Hamiltonian is not a principle limitation 
of the present method but is adequate for the concrete device geometry 
that we will study in detail in this paper. The eigenstates of 
$H_{2}$ are linear combinations of the ground states 
$\left\vert 0\right\rangle $ and $\left\vert 1\right\rangle $ of the two 
isolated quantum dots,
\begin{align}
\left\vert B\right\rangle &=h_{B0}\left\vert 0\right\rangle +h_{B1}\left\vert
1\right\rangle , \label{BondingState}\\
\left\vert A\right\rangle &=h_{A0}\left\vert 0\right\rangle 
+h_{A1}\left\vert
1\right\rangle , \label{AntibondingState}
\end{align}
with real valued coefficients $h_{YJ}$ ($Y=A,B$, $J=0,1$). Here, 
$\left\vert B\right\rangle $ is the bonding eigenstate and 
$\left\vert A\right\rangle $ the anti-bonding eigenstate. The 
corresponding eigenenergies are
\begin{align}
E_{B}&=-\frac{1}{2}\sqrt{t^{2}+\Delta^{2}},\\
E_{A}&=\frac{1}{2}\sqrt{t^{2}+\Delta^{2}}.
\end{align}
For vanishing $t$, the eigenstates $\left\vert B\right\rangle $ and 
$\left\vert A\right\rangle $ reduce to $\left\vert 0\right\rangle $ 
and $\left\vert 1\right\rangle $.

\subsection{Interaction matrix elements}

The Coulomb interaction $V_{12}$ between particle $1$ and particle $2$ 
yields the following expression for the interaction matrix elements 
\begin{align}
\langle\alpha_{i}Y| & V_{12}|\alpha_{j}Z\rangle \notag\\
&=\frac{e^{2}}{4\pi\varepsilon
_{r}\varepsilon_{0}}\int\mathrm{d}^{3}x\,\langle\alpha_{i}|\mathbf{x}
\rangle\langle\mathbf{x}|\alpha_{j}\rangle\sum_{J=0,1}\frac{h_{YJ}h_{ZJ}
}{\left\vert \mathbf{x}-\mathbf{x}_{J}\right\vert },
\end{align}
where $\left\vert \alpha_{i} Y \right\rangle$ is the product state of
particle $1$ state $\left\vert \alpha_{i}\right\rangle$, and particle
$2$ state $\left\vert Y\right\rangle$.

\subsection{Ballistic transport through a 
system of two interacting particles}

In this second step, we develop a Green's function method by 
extending the CBR method\cite{MSV2003, MVSZV2005} to the case of 
an open device that describes a system of two interacting particles. 
We stress that we will consider the two particles as distinguishable.
This is not an approximation for the proposed two-qubit devices
but comes directly from the system geometry. Indeed, we choose the 
QWRs and the double dot to be well separated from each other 
so that no significant 
tunneling between the two structures can occur. In other words,
the two wave functions -- namely the bound state in the double dot 
and the scattering state in the QWRs -- never overlap. This 
leads to the distinguishability of the two particles based on 
their spatial localization. In addition, we assume that the 
interaction is negligible outside the device, due to screening in 
the contacts and/or due to barriers. In the following, the term 
\textit{device} denotes a finite three-dimensional region that is 
coupled to reservoirs by an arbitrary number of leads. The device 
may be under applied bias and contain some spatially varying 
potential. The total (two-particle) Hamiltonian of the system, 
including the device and the leads, can be written in symbolic 
matrix form
\begin{equation}
H_{\mathrm{tot}}=
\begin{pmatrix}
H_{1}^{L} & 0 & 0 & W_{1}\\
0 & \ddots & 0 & \vdots\\
0 & 0 & H_{N_{L}}^{L} & W_{N_{L}}\\
W_{1}^{\dagger} & \cdots & W_{N_{L}}^{\dagger} & H^{0}
\end{pmatrix}
, \label{CBRTotalHamiltonian}
\end{equation}
where $H_{\lambda}^{L}$ represent the Hamiltonian of lead $\lambda$, the 
Hamiltonian $H^{0}$ corresponds to the device region, and $W_{\lambda}$ 
is the coupling between the device and this lead 
($\lambda=1,\dots,N_{L}$). $H^{0}$ is 
composed of the single-particle Hamiltonians $H_{1}$ and $H_{2}$, 
corresponding to particle 1 and 2, respectively, and the 
interaction term $V_{12}$,
\begin{equation}
H^{0}=H_{1}+H_{2}+V_{12}.
\end{equation}
The leads (acting as reservoirs) are semi-infinite and therefore, the 
total Hamiltonian $H_{\mathrm{tot}}$ is infinite-dimensional. This 
infinite-dimensional Hamiltonian can be reduced to a non-Hermitian 
finite-dimensional 
Hamiltonian $H=H^{0}+\Sigma$ that describes the \textit{open device}
exactly\cite{DattaBook}. 
In this formulation, the influence of the leads is included through a 
finite-dimensional operator $\Sigma=\Sigma_{1} +\cdots+\Sigma_{N_{L}}$. 
This is the sum of the complex contact self-energies 
$\Sigma_{\lambda}$ that are nonzero only in 
the \textit{contact regions} where the lead $\lambda$ adjoins to the 
device. The self-energies couple only particle 1 to the leads 
since particle 2 is assumed to be bound. The 
Hermitian Hamiltonan $H^{0}$ represents the \textit{decoupled 
device}, i.e.\ the device with no coupling to the leads. 
$H_{1}$ as well as $H_{2}$ are 
Hamiltonians of the closed device. In the ballistic case, 
all observables of interest such as transmission 
functions and the current can be calculated from the retarded Green's 
function $G^{R}$ of the open device. This is defined by
\begin{equation}
G^{R}=\left(  E-H\right)^{-1}=(E-H_{1}-H_{2}-V_{12}-\Sigma)^{-1}~.
\end{equation}
The retarded Green's function can be obtained from the Green's function $G^{0}$ 
of the decoupled device by the Dyson equation
\begin{equation}
G^{R}=\left(  1-G^{0}\Sigma\right)^{-1}G^{0}, \label{DysonEquation}
\end{equation}
which in turn can be evaluated from its spectral representation
\begin{equation}
G^{0}=\sum_{n}\frac{\left\vert n\right\rangle \left\langle n\right\vert
}{E-\varepsilon_{n}+i\eta},\qquad\eta\rightarrow0^{+},
\label{CBRSpectralRepresentation}
\end{equation}
\begin{equation}
H^{0}\left\vert n\right\rangle =\varepsilon_{n}\left\vert n\right\rangle .
\end{equation}
The direct evaluation of $G^{R}$ according to Eq.~(\ref{DysonEquation}) 
requires the inversion of a huge matrix that is proportional to the 
number of grid points $N_{D}$ of the device. By contrast, the CBR 
method allows one to drastically reduce this effort by utilizing the 
following exact properties of $G^{R}$ that remain valid in the present 
many-particle case. We decompose $G^{R}$ into four blocks,
\begin{equation}
G^{R}=
\begin{pmatrix}
G_{C}^{R} & G_{CD}^{R}\\
G_{DC}^{R} & G_{D}^{R}
\end{pmatrix}
, \label{CBRDecompositionGR}
\end{equation}
where $G_{C}^{R}$ is a matrix proportional to the number $N_{C}$ of 
contact grid points and is called the \textit{contact block}. 
Note that $N_{C}\ll N_{D}$. For 
the contact block $G_{C}^{R}$, the following equation holds:
\begin{equation}
G_{C}^{R}=\left(  1_{C}-G_{C}^{0}\Sigma_{C}\right)  ^{-1}G_{C}^{0},
\label{DysonEquationContactBlock}
\end{equation}
where $1_{C}$, $G_{C}^{0}$, and $\Sigma_{C}$ are the corresponding contact 
blocks of the unity matrix, $G^{0}$, and $\Sigma$, respectively. 
The key point is that this is a linear matrix equation of the order 
of $N_{C}$, since $\Sigma$ 
is nonzero only in the contact region. As a consequence, this contact 
block is also sufficient to determine the transmission functions through 
the device. The ballistic current can then be calculated from the 
transmission functions according to the Landauer-B\"{u}ttiker 
formalism.~\cite{But1986} In the following, we represent $G_{C}^{R}$ in 
the mixed basis of position eigenstates of particle 
$1$ and energy eigenstates of particle $2$,
\begin{equation}
G_{C}^{R}(\mathbf{x}_{i},\mathbf{x}_{j},Y,Z)=\sum_{\alpha_{i},\alpha_{j}
}\langle\mathbf{x}_{i}|\alpha_{i}\rangle\langle\alpha_{i}Y|G_{C}^{R}
|\alpha_{j}Z\rangle\langle\alpha_{j}|\mathbf{x}_{j}\rangle.
\label{GRCMixedBasis}
\end{equation}
In this form, $G_{C}^{R}$ represents the probability amplitude for the 
propagation of particle $1$ from position $\mathbf{x}_{j}$ to position 
$\mathbf{x}_{i}$, accompanied by a transition of particle $2$ from 
eigenstate $\left\vert Z\right\rangle $ to $\left\vert Y\right\rangle $. 
For $\lambda \neq\lambda^{\prime}$, the expression
\begin{eqnarray}
T_{\lambda\lambda^{\prime}}^{Y,Z}\left(  E\right) &=&\mathrm{Tr}_1\,
\Gamma_{C}^{\lambda}\left(  G_{C}^{R}\right)  \Gamma_{C}
^{\lambda^{\prime}}\left(  G_{C}^{R\dagger}\right),
\label{CBRConditionalTransmission}  \\
\Gamma_{C}^{\lambda} 
&=& i\left(  \Sigma_{C}^{\lambda}-\Sigma_{C}^{\lambda\dagger
}\right).
\end{eqnarray}
therefore yields the probability for the transmission of particle $1$ 
from lead $\lambda^{\prime}$ to lead $\lambda$ under the condition, 
that initially particle $2$ is in state $\left\vert Z\right\rangle $ and 
ends up in state $\left\vert Y\right\rangle $. Note that the trace in 
Eq.~(\ref{CBRConditionalTransmission}) is taken only with respect to the 
position variables of particle $1$. If the final state of particle $2$ 
is not measured, the corresponding transmission probability is obtained 
by summing over all final states of particle $2$
\begin{equation}
T_{\lambda\lambda^{\prime}}^{Z}=\sum_{Y}T_{\lambda\lambda^{\prime}}^{Y,Z}.
\end{equation}

In our concrete calculations (see Sec.~\ref{results}), we found it most 
efficient to first determine the eigenstates of the single-particle 
Hamiltonians $H_{1}$ and $H_{2}$ and then diagonalize the interaction 
$V_{12}$ in the product basis of these single-particle eigenstates. This 
procedure allows us to take advantage of another built-in efficiency of 
the CBR method, namely the fact that only energetically low lying 
eigenstates of the closed device couple to the incoming and outgoing 
lead states.~\cite{MSV2003, MVSZV2005} Consequently, it 
suffices to take into account only a reduced set of eigenstates of the 
single-particle Hamiltonians $H_{1}$ and $H_{2}$ without affecting the 
transmission results noticeably.

Another quantity that can be readily obtained from the retarded Green's 
function $G^{R}$ is the charge density of the current carrying states 
of particle $1$. This requires one to know not only the contact 
block $G_{C}^{R}$ but also the submatrix $G_{DC}^{R}$. However, the 
calculation of the submatrix $G_{DC}^{R}$ only requires one to evaluate 
the inverse of the small matrix $1_{C}-\Sigma_{C}G_{C}^{0}$,
\begin{equation}
G_{DC}^{R}=G_{DC}^{0}\left(  1_{C}-G_{C}^{0}\Sigma_{C}\right)  ^{-1},
\end{equation}
where the matrix $G_{DC}^{0}$ corresponds to $G_{DC}^{R}$ in the open 
device.~\cite{MVSZV2005} According to Eq.~(\ref{GRCMixedBasis}), the 
latter Green's function reads in the mixed basis
\begin{equation}
G_{DC}^{R}(\mathbf{x}_{i},\mathbf{x}_{j},Y,Z)=
\sum_{\alpha_{i},\alpha_{j}
}\langle\mathbf{x}_{i}|\alpha_{i}\rangle\langle\alpha_{i}Y|G_{DC}^{R}
|\alpha_{j}Z\rangle\langle\alpha_{j}|\mathbf{x}_{j}\rangle.
\end{equation}
The total charge density of the current carrying states of particle $1$ 
is consequently given by
\begin{equation}
n\left(  \mathbf{x}_{j}\right)  =
\sum_{Z}n^{Z}\left(  \mathbf{x}_{j}\right)  ,
\end{equation}
where the contribution
\begin{align}
n^{Z}\left(  \mathbf{x}_{j}\right)   &  =
\frac{1}{2\pi}\sum_{\lambda}\sum
_{Y}\int\mathrm{d}E\,\Xi_{\lambda}^{Y,Z}\left(  E\right)  f\left(
\frac{E-E_{\mathrm{F}}^{\lambda}}{k_{\mathrm{B}}T}\right)
,\label{ProjectedDensity} \\
\Xi_{\lambda}^{Y,Z}\left(  E\right)   &  =\sum_{\mathbf{x}_{j}}G_{DC}
^{R}(\mathbf{x}_{i},\mathbf{x}_{j},Y,Z)  \notag \\
& \times \Gamma_{C}^{\lambda}(\mathbf{x}
_{j},\mathbf{x}_{j})\left(  G_{DC}^{R}(\mathbf{x}_{i},\mathbf{x}
_{j},Y,Z)\right)  ^{\dagger}
\end{align}
results from the projection of the total charge density onto the 
eigenstate $\left\vert Z\right\rangle $ of the second particle. Here, 
$E_{\mathrm{F} }^{\lambda}$ denotes the Fermi level in lead $\lambda$.

\section{Numerical details}  \label{numericaldetails}

\subsection{Electronic structure of particle 1}

Equations~(\ref{SchroedingerEquation}) and 
(\ref{NonlinearPoissonEquation}) are coupled partial differential 
equations in position space that we discretize and map onto a nonuniform 
tensor product grid. For the discretization of the 
all-electric Mach-Zehnder interferometer, a total 
of $6\times10^{6}$ grid points have been used. We employ box integration 
finite differences in order to ensure 
flux conservation across boundaries with different material parameters. 
For the solution of the nonlinear system that results from the 
discretization of the nonlinear Poisson 
equation (\ref{NonlinearPoissonEquation}), Newton's method with inexact 
line search is invoked. This remaps the problem into a 
sequence of linear solutions. We use the Dupont-Kendall-Rachford 
preconditioned conjugate gradient method for the solution of the 
resulting linear systems.~\cite{BDDRV2000, DKR1968} The discretization 
of the Schr\"{o}dinger equation (\ref{SchroedingerEquation}) results 
in a large matrix eigenvalue system. However, since the occupation of 
electron states falls off exponentially with increasing energy distance 
from the Fermi level, only about $250$ quantum states suffice for a 
converged calculation of the electron density. This allows 
for the use of the iterative Arnoldi method that is implemented 
in the published ARPACK libraries.~\cite{ARPACK} We found that the 
calculation of eigenvalues and eigenstates can be accelerated by an 
order of magnitude by invoking a Chebyshev polynomial based spectral 
transformation that provides an efficient preconditioning of the 
linear system.~\cite{AZABSMV2006} Finally, in order to solve the 
coupled Poisson-Schr\"{o}dinger system, we employ an approximate 
quantum charge density inside of Poisson's equation in order to 
estimate the dependence of the density on the potential through 
Schr\"{o}dinger's equation. This corrector-predictor scheme reduces 
the number of required diagonalizations and accelerates the 
convergence of the coupled system significantly.~\cite{TGPR1997}

Equation~(\ref{NonlinearPoissonEquation}) requires the specification 
of boundary conditions in order to obtain unique solutions. 
For the potential in the Poisson equation, we generally use von 
Neumann boundary conditions. Dirichlet boundary conditions are 
employed at Schottky contacts and at the GaAs/air interfaces. In 
the former case, the electrostatic potential at the boundary is set 
equal to the Schottky barrier height, i.e.\ the difference between 
the conduction band edge and the quasi-Fermi level. At the GaAs/air 
interface, on the other hand, the electrostatic potential is determined 
by the experimentally known Fermi level pinning. The Schr\"{o}dinger 
equation for the open quantum system is solved in 
terms of the CBR method that has been described 
in detail elsewhere.~\cite{MVSZV2005} For the calculation of the 
transmission probabilities, we have taken into account the lowest 
$250$ eigenstates for the determination of the retarded Green's 
function of the closed device and the 
lowest $5$ propagating modes in each of the four leads. The 
self-consistent solution of Eqs.~(\ref{SchroedingerEquation}) and 
(\ref{NonlinearPoissonEquation}), for given split gates voltages, takes 
approximately $2$ days on state-of-the-art PC hardware.  Converged 
self-consistent results require typically $12$ iterations in the two 
sets of equations.

\subsection{Entangled Mach-Zehnder double quantum dot device}

Our novel method for the calculation of the ballistic transport 
properties through a system of two interacting particles requires the 
diagonalization of the two-particle Hamiltonian including the 
interaction. This has to be done once for every bias point in the 
$I$-$V$ characteristics. The diagonalization is 
performed in the set of product states formed by subsets of the 
single-particle eigenstates of the two non-interacting particles. 
We found that a subset of $100$ 
lowest eigenstates of particle $1$ is sufficient to obtain convergent 
transmission functions in the meV energy range around the Fermi level 
that is relevant for this device.

\section{Analytical model of the two-qubit device}  
\label{analyticalmodel}

\begin{figure}
\begin{center}
\includegraphics{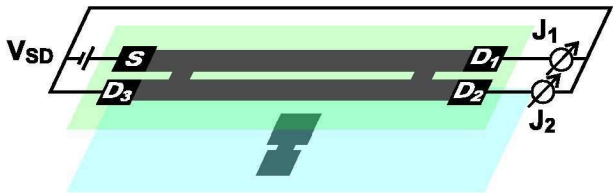}
\caption{(Color online) Schematic view of the proposed 
quantum transport device. The device is realized by two stacked 
GaAs/AlGaAs 2DEGs. The top 2DEG (green) is depleted by external gates 
to form a double-QWR based Mach-Zehnder interferometer. The bottom 
2DEG (blue) contains two electrostatically defined, coupled quantum 
dots. The larger one is centered beneath and in between the 
QWRs. There is a bias voltage $V_{\mathrm{SD}}$ applied between the 
source ($S$) and the three drain contacts ($D_1 - D_3$) that leads to 
currents ($J_1, J_2$) that are labeled accordingly. There are 
additional gates acting on the device that are shown in the following 
Fig.~\ref{HeterostructureAndMetalGates}. 
For sake of clarity, the figure is not drawn to scale.
}
\label{SchematicDevice3D}
\end{center}
\end{figure}

Before presenting and discussing the numerical results, we develop 
a simple analytical model of our two-qubit Mach-Zehnder double-dot 
system in the present section. The device is shown schematically 
in Fig.~\ref{SchematicDevice3D}. The purpose of this procedure is to 
be able to better grasp the physics behind the generation of 
entanglement before we present the numerical results of the fully
three-dimensional self-consistent calculations in the next section.

The good agreement between the two approaches will be a further
confirmation of the correct functioning of the device: The results
that we obtain for the entangled Mach-Zehnder double quantum dot
device are indeed the result of a two-qubit quantum
operation.  Obviously, the results of the two methods will only be in
qualitative agreement with one another, since the analytical model 
developed in the this section cannot account for the full complexity 
of the realistic device.

We describe the device consisting of a Mach-Zehnder interferometer 
coupled to a double quantum dot, by two coupled qubits, W and D, 
respectively. The basis states 
$\left\vert 0\right\rangle^{\mathrm{W}}$ and 
$\left\vert 1\right\rangle ^{\mathrm{W}}$ of qubit W are defined by 
an electron that propagates through either of the two QWRs.
The basis states 
$\left\vert 0\right\rangle^{\mathrm{D}}$ and 
$\left\vert 1\right\rangle ^{\mathrm{D}}$ of qubit D, on the other 
hand, are defined by the ground states of the isolated
quantum dots.  Quantum gates can be described by unitary operators that
map an initial state $\left\vert \Phi_{\mathrm{in}}\right\rangle $
onto a corresponding final state 
$\left\vert \Phi_{\mathrm{fi}}\right\rangle$.  The unitary 
operator $\hat{T}$ that
represents our two-qubit quantum transport device can be decomposed
into a sequence of unitary operators of less complex single- and
two-qubit gates as follows\cite{BBBJR2002}:
\begin{equation}
\hat{T}=\hat{R}\cdot\hat{P}\cdot\hat{V}\cdot\hat{R}. 
\label{SequenceOfGates}
\end{equation}
The operators $\hat{R}$ and $\hat{P}$ represent single-qubit gates 
that act solely on qubit W.
In particular, $\hat{R}$ represents a $\pi/2$ rotation gate and
describes a coupling window, whereas $\hat{P}$ represents a phase gate
that adds a phase difference to the two components of the qubit. The 
latter can be realized through an additional split gate located on 
one of the QWRs, as explained in Sec.~\ref{results}.
The sequence $\hat{R}\cdot\hat{P}\cdot\hat{R}$ therefore corresponds to 
a Mach-Zehnder interferometer. The operator $\hat{V}$ represents a 
two-qubit gate and describes 
the coupling of the two qubits D and W due to their interaction. This 
decomposition allows us to specify the action of the single-qubit gates 
$\hat{R}$ and $\hat{P}$ directly in terms of unitary operators.

We proceed by describing the entangled Mach-Zehnder double dot device 
as a stationary quantum-mechanical scattering problem. To this end, 
we represent the two electrons in the pair of QWRs and the double 
quantum dot by the following 8 basis states for given total energy 
$E$ and a given position $x$,
\begin{equation}
\psi_{\pm k_{J},\sigma J}\left(  x\right)  =\exp\left(  \pm ik_{J}x\right)
\left\vert \sigma\right\rangle ^{\mathrm{W}}\left\vert J\right\rangle
^{\mathrm{D}}.\qquad (\sigma,J=0,1)
\end{equation}
These states describe the two electrons as follows: The first electron 
propagates along the wire axis $x$ in either of the two states 
$\left\vert 0\right\rangle ^{\mathrm{W}}$ and 
$\left\vert 1\right\rangle ^{\mathrm{W}}$, 
corresponding to the two QWRs. The electron 
momentum at position $x$ is 
denoted by $k_{J}$. The second electron occupies one of the two quantum 
dots, represented by states 
$\left\vert 0\right\rangle ^{\mathrm{D}}$ and $\left\vert 
1\right\rangle ^{\mathrm{D}}$. In this representation, the operators in
Eq.~(\ref{SequenceOfGates}) correspond to $8\times8$ transfer matrices,
\begin{equation}
T=T_{\hat{R}}\cdot T_{\hat{P}}\cdot T_{\hat{V}}\cdot T_{\hat{R}}\text{ .}
\end{equation}
This will allow us to finally determine the transmission function 
through the device. As a first step, we derive the transfer matrix 
$T_{\hat{V}}$ of the two-qubit gate $\hat{V}$ by solving a 
two-particle Schr\"{o}dinger equation. 
The non-interacting two-particle Hamiltonian that describes the 
dynamics of the system is given by
\begin{equation}
H_{0}=\frac{p_{x}^{2}}{2m}+H_2.
\end{equation}
Here, $H_2$ is the model Hamiltonian that describes a pair of 
tunneling coupled quantum dots in terms of the bare splitting 
$\Delta$ and the tunneling coupling $t$ in accordance with 
Eq.~(\ref{ModelHamiltonianQuantumDots}).  For the Coulomb interaction 
between the two electrons, we make the following ansatz
\begin{equation}
V\left(  x\right)  =U\left\vert 1\right\rangle ^{\mathrm{W}}\left\vert
1\right\rangle ^{\mathrm{D}}\left\langle 1\right\vert ^{\mathrm{W}
}\left\langle 1\right\vert ^{\mathrm{D}}\theta\left(  x\right) 
\theta\left(L-x\right)  .
\end{equation}
This describes a localized interaction with interaction strength $U$ 
that is nonzero only for the two-particle state 
$\left\vert 1\right\rangle ^{\mathrm{W} 
}\left\vert 1\right\rangle ^{\mathrm{D}}$ but zero for all other states. 
In addition, the interaction vanishes
if the position of the QWR electron lies outside the spatial 
interval from $x=0$ to $x=L$. The 
resulting quantum mechanical scattering problem
\begin{equation}
H=H_{0}+V\left(  x\right)  ,
\end{equation}
can be solved analytically for the $8\times8$ transfer 
matrix $T_{\hat{V}} (E)$. 
Since $V\left(  x\right)  $ only acts on the state 
$\left\vert 1\right\rangle^{\mathrm{W}}\left\vert 1\right\rangle 
^{\mathrm{D}}$, 
$T_{\hat{V}}$ has the following form,
\begin{equation}
T_{\hat{V}}=
\begin{pmatrix}
\mathbf{1}_{4\times4} & 0\\
0 & \mathbf{\tau}
\end{pmatrix}
,
\end{equation}
where $\mathbf{\tau}$ is a $4\times4$ submatrix that connects the states 
$\sigma=1$, i.e.\ states that contain 
$\left\vert 1\right\rangle ^{\mathrm{W}} $. 
In order to determine this submatrix, we split the $x$-axis into 3 regions 
$x<0$, $0<x<L$, and $x>L$. Within each region, the Hamiltonian $H$ can be 
diagonalized. Requiring continuity and differentiability, we obtain the 
following explicit expression for the submatrix $\mathbf{\tau}$. In this 
expression, we have transformed the basis by using 
the bonding and antibonding 
dot states $\left\vert A\right\rangle ^{\mathrm{D}}$ and $\left\vert 
B\right\rangle ^{\mathrm{D}}$ from 
Eqs.~(\ref{BondingState}), (\ref{AntibondingState}) 
instead of the states $\left\vert 0\right\rangle ^{\mathrm{D}}$ and 
$\left\vert 1\right\rangle ^{\mathrm{D}}$. In this basis, we obtain 
$\mathbf{\tau}=\mathbf{\kappa}_{L}^{-1}\mathbf{\mu}_{L}\mathbf{\kappa 
}_{0}^{-1}\mathbf{\mu}_{0}$ where the submatrices $\mu,\kappa$ read 
for $x_1=0$, $x_2=L$
\begin{equation}
\mathbf{\mu}_{x_i}=
\begin{pmatrix}
\nu_{x_i}\left(  k_{B}\right)  & 0\\
0 & \nu_{x_i}\left(  k_{A}\right)
\end{pmatrix}
, \label{TransferMatrix1}
\end{equation}
\begin{equation}
\mathbf{\kappa}_{x_i}=
\begin{pmatrix}
\alpha_{B}\nu_{x_i}\left(  q_{B}\right)  & 
\alpha_{A}\nu_{x_i}\left(  q_{A}\right)
\\
\beta_{B}\nu_{x_i}\left(  q_{B}\right)  & 
\beta_{A}\nu_{x_i}\left(  q_{A}\right)
\end{pmatrix}
, \label{TransferMatrix2}
\end{equation}
where $i=1,2$. The $2\times2$ matrix $\nu_{x_i}\left(k\right)$ is 
defined as follows:
\begin{equation}
\nu_{x_i}\left(  k\right)  =
\begin{pmatrix}
\exp\left(  ikx_i\right)  & \exp\left(  -ikx_i\right) \\
ik\exp\left(  ikx_i\right)  & -ik\exp\left(  -ikx_i\right)
\end{pmatrix}
.
\end{equation}
The coefficients $\alpha_{Y}$, $\beta_{Y}$ and the wave vectors $k_{Y}$,
$q_{Y}$ ($Y=A,B$) are given by ($t,U>0$)
\begin{equation}
\alpha_{Y}=\frac{tU}{\sqrt{\left(  4E_{Y}\left(  E_{Y}+\varepsilon\right)
+\Delta U\right)  ^{2}+t^{2}U^{2}}},
\end{equation}
\begin{equation}
\beta_{Y}=\frac{4E_{Y}\left(  E_{Y}+\varepsilon\right)  +\Delta U}
{\sqrt{\left(  4E_{Y}\left(  E_{Y}+\varepsilon\right)  +\Delta U\right)
^{2}+t^{2}U^{2}}}.
\end{equation}
\begin{equation}
k_{Y}=\frac{1}{\hbar}\sqrt{2m\left(  E-E_{Y}\right)  },
\end{equation}
\begin{equation}
q_{Y}=\frac{1}{\hbar}\sqrt{2m\left( E-\frac{1}{2}U\mp\varepsilon\right) },
\end{equation}
where $\varepsilon=\frac{1}{2}\sqrt{t^{2}+\left(  U+\Delta\right)^{2}}$.
Note that the electron wave vectors $k_{Y}$ arise from the region $x<0$ 
and from $x>L$, whereas $q_{Y}$ arises from the central region.

The next step is to determine the transfer matrices $T_{\hat{R}}$ and
$T_{\hat{P}}$ from the following representations of ideal rotation and 
phase gates,
\begin{equation}  \label{eqRtransf}
\hat{R}=
\begin{pmatrix}
\cos\left(  \theta/2\right)  & -\sin\left(  \theta/2\right) \\
\sin\left(  \theta/2\right)  & \cos\left(  \theta/2\right)
\end{pmatrix}
,\qquad\theta=\pi/2,
\end{equation}
\begin{equation}  \label{eqPtransf}
\hat{P}=
\begin{pmatrix}
e^{i\phi/2} & 0\\
0 & e^{-i\phi/2}
\end{pmatrix}
,
\end{equation}
where $\phi$ is the phase gate angle. We assume that these gates do not
introduce back-scattering. In this case, we can apply $\hat{R}$ and 
$\hat{P}$ separately for both directions of propagation. This leads 
to unitary matrices
$T_{\hat{R}},T_{\hat{P}}$ with the following nonvanishing elements
\begin{align}
(T_{\hat{R}})_{j,j}  &  =\cos\left(  \theta/2\right), \quad(T_{\hat{R}}
)_{i,i+4}=(-1)^{i}\sin(\theta/2)\text{ },\notag\\
(T_{\hat{R}})_{i+4,i} & =(-1)^{i+1}
\sin(\theta/2),\quad(i=1,4;j=1,8)\\
\left(  T_{\hat{P}}\right)  _{j,j}  &  =\exp\left[  (-1)^{j+1}\frac{\phi}
{2}\right]\quad(j=1,8).
\end{align}
This completes the calculation of the total transfer matrix $T$. 
The final step is to determine the transmission function which requires 
us to set an initial 
condition for the scattering problem. We consider the situation where the 
propagating electron enters the pair of QWRs from the left in state 
$\left\vert 0\right\rangle ^{\mathrm{W}}$. We further assume that the 
electron in the double quantum dot lies initially in the binding state 
$\left\vert 
B\right\rangle ^{\mathrm{D}}$. The eight final state coefficients 
represent the transmission amplitudes into the four states 
$\left\vert \sigma\right\rangle^{\mathrm{W}}\left\vert Y\right\rangle 
^{\mathrm{D}}$ ($\sigma=0,1$, $Y=A,B$) 
for each of the two asymptotic propagation directions (to the right 
or to the left). We denote these final state coefficients by 
$E_{\sigma,Y}^{r}$ and $E_{\sigma,Y}^{l}$, respectively. By 
construction, the two rotation gates and the phase gate do not introduce 
any back-scattering. The amount of back-scattering 
due to the interaction $V\left(  x\right)  $ depends on the ratio of 
the interaction strength $U$ and the length $L$ of the interaction 
region and can be made arbitrarily small. Thus, the transmission 
amplitudes $E_{\sigma,Y}^{l}$ are negligible. In addition, we only need 
the transmission probabilities $T_{1}$ and $T_{2}$ that describe the 
total probability of an electron to arrive at 
$\left\vert 0\right\rangle ^{\mathrm{W}}$ or 
$\left\vert 1\right\rangle ^{\mathrm{W}}$, irrespective of the state of 
the double dot electron,
\begin{align}
T_{1}  &  =\left\vert E_{0,A}^{r}\right\vert ^{2}+\left\vert E_{0,B}
^{r}\right\vert ^{2}, \label{transmission1} \\
T_{2}  &  =\left\vert E_{1,A}^{r}\right\vert ^{2}+\left\vert E_{1,B}
^{r}\right\vert ^{2}. \label{transmission2}
\end{align}

The resulting transmission probabilities $T_{1}$ and $T_{2}$ will be 
presented in the following section for different values of the 
tunneling coupling. They will be compared with the numerical results
obtained with the realistic model described in Sec.~\ref{method}.
Indeed, the above comparison will show
that the two models are able to reproduce the basic physics of 
the device and that the device is able to produce
an entangled state of the two qubits.

\section{Results and Discussion}  
\label{results}

The method presented in the Sec.~\ref{method} allows us to calculate
the ballistic current through entangled two-particle systems.
Concretely, we first present results for a double-QWR based
Mach-Zehnder interferometer and show that this interferometer can be
employed as a fully controllable single-qubit gate.  Based on this
system, we then propose a novel two-qubit quantum transport device
that allows the controlled generation and measurement of entanglement
between the QWR qubit and the double quantum dot qubit.  The
measurement involves only the DC $I$-$V$ characteristics, no higher
order current correlations or magnetic fields are required to detect
entanglement.  We would like to emphasize that the subject of this
paper is to show that these devices are well suited to prepare
arbitrary single-qubit and entangled two-qubit states.  Actual quantum
computing applications, however, require additionally the performance
of these operations in a time-resolved fashion with single electrons.

A schematic view of the proposed two-qubit device is shown in 
Fig.~\ref{SchematicDevice3D}. It is based on a GaAs/AlGaAs 
heterostructure that consists of two vertically stacked 2DEGs.
The 2DEGs are locally depleted by the application of negative voltages 
to appropriate metal top and bottom gates, that are not 
shown in this figure.  For suitable gate voltages, the top gates 
create a pair of parallel QWRs in the upper 2DEG that are connected to 
each other by two coupling windows. For specific energies, these 
coupling windows act as beam-splitters and 
the QWR network behaves as an electronic Mach-Zehnder interferometer. 
In contrast to electronic Mach-Zehnder interferometers based on quantum 
Hall edge channels,\cite{JCSHMS2003} no magnetic fields are employed. 
We therefore use the term \textit{all-electric Mach-Zehnder 
interferometer} for this kind of 
device. We note that the relative phase shift between the electronic wave 
function in the two QWRs is controlled electrostatically by 
phase gates. These are additional top gates, but may be combined with 
the gates that are used to define the QWRs in the first place. The 
lower 2DEG contains the two electrostatically defined tunneling-coupled 
single-electron quantum dots. An electron in the upper 2DEG couples to 
the electron in the double quantum dot in the lower 2DEG by means of 
the Coulomb interaction. We have designed the entire device so that 
the consequences of entanglement become markedly evident in the 
observables, and we refer to it 
as \textit{Entangled Mach-Zehnder Double Quantum Dot Device}.

The Mach-Zehnder interferometer is operated under ballistic transport 
conditions. By applying a small DC bias voltage $V_{\mathrm{SD}}$ 
between the upper left source $S$ and the remaining three drain contacts 
$D_{1} ,D_{2},D_{3}$, the device can be used in such a way that 
current flows predominantly from $S$ to $D_{1}$ (denoted by $J_{1}$) 
or to $D_{2}$ (denoted by $J_{2}$), as indicated 
in Fig.~\ref{SchematicDevice3D}.

\subsection{Results: All-electric Mach-Zehnder interferometer (single 
qubit device)}

\begin{figure}
\begin{center}
\includegraphics{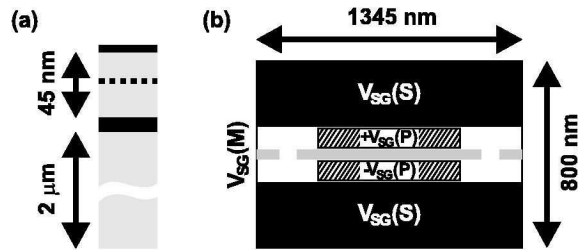}
\caption{(a) Vertical cross section of the GaAs/AlGaAs heterostructure 
with a single 10 nm wide GaAs quantum well that is used for the 
Mach-Zehnder interferometer. GaAs is shown in black, 
Al$_{0.37}$Ga$_{0.63}$As is shown in gray. The GaAs cap layer is 5 nm 
thick. A silicon $\delta$ doping layer with 
a concentration of $2.5\times10^{12}~\mathrm{cm}^{-2}$ 
is located $25~\mathrm{nm}$ below the surface (dashed line). (b) Top 
view of the Mach-Zehnder interferometer that depicts the side gates 
(black) with voltage $V_\mathrm{SG}(\mathrm{S})$ and the 
mid gates (gray) with voltage $V_\mathrm{SG}(\mathrm{M})$ that define 
the two QWRs and the coupling windows between them. On top of the 
wires, there are two additional phase gates (hatched) 
at a bias of $\pm V_\mathrm{SG}(\mathrm{P})$.}
\label{HeterostructureAndMetalGates}
\end{center}
\end{figure}

First, we turn to the $I$-$V$ characteristics of the single-electron 
device, i.e.\ the all-electric Mach-Zehnder interferometer alone. 
The calculations have been based on the GaAs/AlGaAs heterostructure that 
is depicted schematically in 
Fig.~\ref{HeterostructureAndMetalGates}. It consists of a 
$5~\mathrm{nm}$ thick GaAs cap layer [see vertical cross section in 
Fig.~\ref{HeterostructureAndMetalGates}~(a)], followed by a 
$45~\mathrm{nm}$ barrier of Al$_{0.37}$Ga$_{0.63}$As, and a 
$10~\mathrm{nm}$ GaAs 2DEG layer. Beneath, there lies a 
$2~\mathrm{\mu m}$ Al$_{0.37}$Ga$_{0.63}$As substrate. A 
silicon $\delta$ doping layer with a concentration of 
$2.5\times10^{12}~\mathrm{cm}^{-2}$ is located $25~\mathrm{nm}$ below 
the surface. The electron sheet density in the 2DEG layer has been 
calculated to be $2.4\times10^{11}~\mathrm{cm}^{-2}$ for the ungated 
sample at a temperature of $4~\mathrm{K}$.

Figure~\ref{HeterostructureAndMetalGates}~(b) shows a top view of the 
structure including the gates. The source and drain contacts (cf. 
Fig.~\ref{SchematicDevice3D}) are not shown in this figure. The device 
is $800~\mathrm{nm}$ wide and $1345~\mathrm{nm}$ long. Here we 
distinguish three types of gates: side gates (black), mid gates (gray), 
and phase gates (hatched). Each of the two side gates are, respectively, 
$280~\mathrm{nm}$ wide, $1345~\mathrm{nm}$ long, and are biased at 
$V_{\mathrm{SG}}\left(  \mathrm{S}\right)  =-0.245~\mathrm{V}$ with 
respect to the source. The three mid gates are $40~\mathrm{nm}$ wide and 
$200~\mathrm{nm}$, $800~\mathrm{nm}$, and $200~\mathrm{nm}$ long, 
respectively. We apply a gate voltage of 
$V_{\mathrm{SG}}\left(\mathrm{M}\right) 
=-0.660~\mathrm{V}$ to the mid gates. They define two QWRs (white) 
with a nominal width of $100~\mathrm{nm}$ as well as two coupling 
windows, each with a nominal length of $72.5~\mathrm{nm}$ in the 
direction parallel to the wires. All of these structural parameters 
have been chosen to guarantee optimal device operation, while allowing 
for its fabrication with current technologies. 

\begin{figure}
\begin{center}
\includegraphics{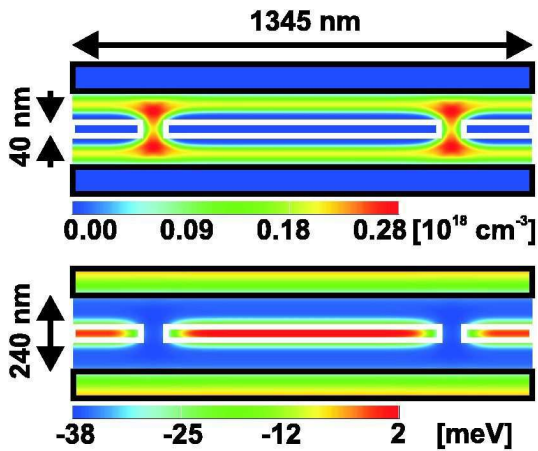}
\caption{(Color online) Top views of the equilibrium electron density 
(upper panel) and electrostatic potential (lower panel) within the 
upper 2DEG. The black and the white frames indicate the position of the 
metal side and mid gates.}
\label{DensityAndPotential}
\end{center}
\end{figure}

The subband spacing of the two lowest subbands in each of these 
QWRs amounts to $3.1~$\textrm{meV}. It is important for the 
interferometer that the Fermi wave length is close to the length of 
the coupling windows. Indeed, our calculations yield a Fermi wave 
length of $\lambda_{\mathrm{F}}=77~\mathrm{nm}$ in the lowest subband. 
In order to control the relative phase of the electron wave function 
in the two QWRs, a small gate voltage 
$+V_{\mathrm{SG}}\left(  \mathrm{P}\right)$ and 
$-V_{\mathrm{SG}}\left(\mathrm{P}\right)  $ is applied to the phase gates, 
respectively. The two phase gates are centered on top of the two QWRs 
and are each $720~\mathrm{nm}$ long and $95~\mathrm{nm}$ wide. A 
separating layer of resist could be used to reliably insulate the phase 
gates from the other gates. Figure~\ref{DensityAndPotential} shows the 
equilibrium charge density (upper panel) and the corresponding 
potential (lower panel) in the 2DEG layer of the Mach-Zehnder 
interferometer. The black and the white framed rectangles indicate the 
position of the metal top gates. The upper panel shows the 2DEG to be 
fully depleted underneath the gates. In addition, it is 
evident from the figure that the QWRs and the coupling windows are 
formed indeed. The wires are strictly one-dimensional 
in the sense that only the lowest subband contributes to the density. 
The lower panel shows the potential barrier that separates the two 
QWRs. The energy scale is chosen so that 
the Fermi level lies at zero $\mathrm{meV}$.

\begin{figure}
\begin{center}
\includegraphics{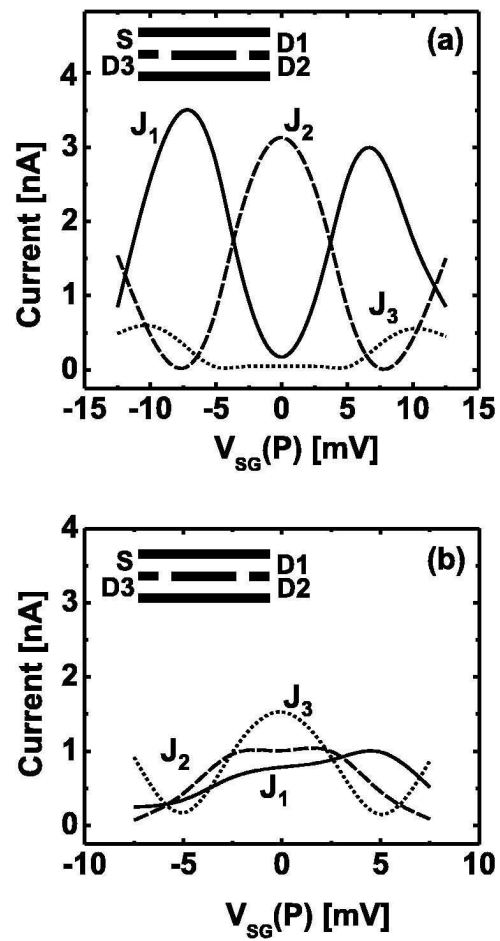}
\caption{(a) Stationary currents $J_{1}$ (solid), $J_{^2}$ (dashed), and 
$J_{^3}$ (dotted) through the Mach-Zehnder interferometer as a function 
of the phase gate voltage $V_\mathrm{SG}(\mathrm{P})$ for a 
nominal length of the coupling windows of $72.5~\mathrm{nm}$. The 
source-drain voltage $V_\mathrm{SD}$ is set to 
$50~\mathrm{\mu V}$. (b) Same as (a), but for a coupling window length 
of $67.5~\mathrm{nm}$.}
\label{CurrentMachZehnder}
\end{center}
\end{figure}

Figure~\ref{CurrentMachZehnder}~(a) shows the currents $J_{1}$ (solid), 
$J_{2}$ (dashed), as well as the back-scattered current 
$J_{3}$ (dotted) that flow from 
the source $S$ to $D_1$, $D_2$, and $D_3$, respectively. The currents 
have been calculated by assuming a DC bias voltage of 
$50~\mathrm{\mu V}$ and a temperature of $30~\mathrm{mK}$. The figure 
shows the dependence of the currents 
on the voltages $+V_{\mathrm{SG}}\left(\mathrm{P}\right)$ and 
$-V_{\mathrm{SG}}\left(\mathrm{P}\right)$ applied to the two phase gates, 
respectively. We first note that the back-current $J_{3}$ is smaller 
than $0.5~\mathrm{nA} $ for the entire relevant range of gate voltages 
and therefore negligible. In contrast to $J_{3}$, the currents 
$J_{1}$ and $J_{2}$ strongly depend on the gate voltage and oscillate 
between $0$ and $4~\mathrm{nA} $ for more than $1.5$ oscillation 
periods. The currents $J_{1}$ and $J_{2}$ are seen to be phase-shifted 
relative to one another by 180 degrees which confirms 
that the device actually behaves as a Mach-Zehnder interferometer. Note 
that this Mach-Zehnder interference pattern becomes damped out once 
the absolute gate 
voltage $|V_{\mathrm{SG}}\left(  \mathrm{P}\right)  |$ 
exceeds $15~\mathrm{mV}$ because this 
increases the amount of back-scattering. Back-scattering can be 
suppressed, however, by increasing the length of the device, because 
the phase shift depends linearly on the length of the phase gate, whereas 
back-scattering is independent of the length but increases with the 
magnitude of $|V_{\mathrm{SG}}\left( \mathrm{P}\right)|$.

For comparison, Fig.~\ref{CurrentMachZehnder}~(b) depicts the calculated 
currents $J_{1}$, $J_{2}$, and $J_{3}$ for the same device but for a 
nominal length of the coupling windows that has been reduced by 
$5~\mathrm{nm} $. The difference in results shows the sensitivity of 
the interference pattern to small changes in geometry, and implies very 
high demands on fabrication precision.

\begin{figure}
\begin{center}
\includegraphics{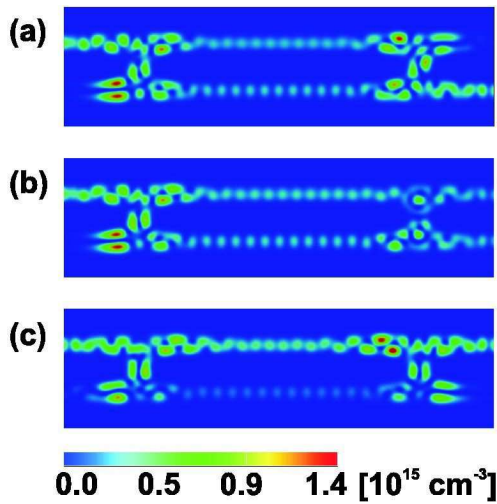}
\caption{(Color online) Charge densities associated with the current 
carrying scattering states for a phase gate gate voltage 
$V_{\mathrm{SG}}(P)$ of (a) 0 meV, (b) -3.6 meV, (c) -7.5 meV. 
In all cases, the source-drain voltage 
$V_\mathrm{SD}$ is set to $50~\mathrm{\mu V}$.}
\label{DensitiesOpenDevice}
\end{center}
\end{figure}

Figure~\ref{DensitiesOpenDevice} shows the charge densities of the 
stationary current carrying states of the Mach-Zehnder interferometer 
for selected phase gate voltages. The three 
Figs.~\ref{DensitiesOpenDevice} (a), (b), and (c) 
correspond to gate voltages of 
$V_{\mathrm{SG}}(\mathrm{P})=0~\mathrm{mV} $, $-3.6 
~\mathrm{mV} $, and $-7.5~\mathrm{mV} $, respectively. The three 
graphs show that the quantum mechanical charge densities change 
predominantly near the drain contacts on the right hand side of the 
device. These changes reflect the redistribution of the total current 
between $J_{1}$ and $J_{2}$. In Fig.~\ref{DensitiesOpenDevice}~(a), the 
charge density is zero near the upper 
right terminal but large near the lower right terminal. This agrees with 
Fig.~\ref{CurrentMachZehnder}~(a) that shows that 
the current $J_{1}$ is minimal ($J_{2}$ maximal) for the corresponding 
gate voltage $V_{\mathrm{SG}}(\mathrm{P})=0 
~\mathrm{mV} $. In Fig.~\ref{DensitiesOpenDevice}~(b), on the other hand, 
the charge density is seen to be almost equal near both right drain 
contacts, again in accord with Fig.~\ref{CurrentMachZehnder}~(a) 
for $V_{\mathrm{SG}}(\mathrm{P})=$ 
$-3.6~\mathrm{mV} $. Finally, the charge density is large near the upper 
right terminal but zero at the lower right terminal in 
Fig.~\ref{DensitiesOpenDevice}~(c), again in correspondence with 
Fig.~\ref{CurrentMachZehnder}~(a).

All of these effects are a consequence of the self-interference of the 
electron wave function. Depending on the relative phase between the two 
QWRs, the interference at the second beam splitter leads to partial 
extinction either near the upper or the lower right terminal.

\subsection{Discussion: Switching characteristics of the
all-electric Mach-Zehnder interferometer}

\begin{figure}
\begin{center}
\includegraphics{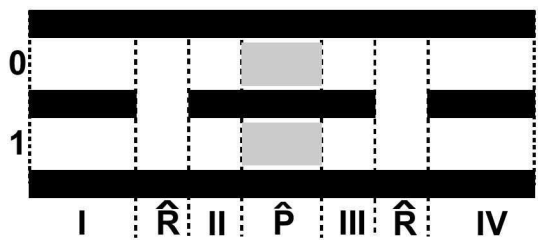}
\caption{Schematic representation of the Mach-Zehnder
interferometer. We have indicated the three quantum-operation
regions, consisting of the two coupling windows $\hat{R}$ and 
one phase gate $\hat{P}$, and four other segments where the qubit
state is well defined (see Table~\ref{tableQWRqubit}).}
\label{QWRqubitDEF}
\end{center}
\end{figure}

The DC transfer characteristics shown in Fig.~\ref{CurrentMachZehnder}~(a) 
exhibits multiple pronounced maxima and minima that can be attributed to 
rotations of a qubit on the Bloch sphere. 
Before entering into the discussion of the results, let us 
give a rigorous definition of the Mach-Zehnder interferometer qubit state.
To this aim, we depict the device schematically in 
Fig.~\ref{QWRqubitDEF} where we show the regions of the two 
coupling windows and the phase gate that perform the 
$\hat{R}$ and $\hat{P}$ transformations
of Eqs.~(\ref{eqRtransf}) and (\ref{eqPtransf}), respectively.
We define the basis states $\left\vert 0\right\rangle $ and
$\left\vert 1\right\rangle $ of the QWR qubit by an electron
scattering state localized in the upper and lower QWR, respectively.
This definition holds only in the regions where the two channels are
well separated, as in the segments \textsf{I}, \textsf{II},
\textsf{III}, and \textsf{IV} of Fig.~\ref{QWRqubitDEF}.  By contrast, 
the qubit state is not well defined within the
window regions $\hat{R}$, where the barrier between the QWRs is 
very small. 
As an example, the approximate qubit states for the four regions 
indicated in
Fig.~\ref{QWRqubitDEF} are included in Table~\ref{tableQWRqubit}, 
and specified for each of the three cases depicted in
Fig.~\ref{DensitiesOpenDevice}.  Indeed, in the calculation of
Fig.~\ref{DensitiesOpenDevice}, the split gate voltages have been
chosen in order to obtain a qubit rotation angle $\theta=\pi/2$ and
phase angles $\phi=0,\pi/2,\pi$ in (a), (b), (c), respectively
[cf. Eqs.~(\ref{eqRtransf}) and (\ref{eqPtransf})].

\begin{table}
\begin{center}
\begin{tabular}{l|llll}
charge  & \multicolumn{4}{|l}{Mach-Zehnder interferometer region} \\ 
density & I & II & III & IV \\ \hline\hline
Fig.~\ref{DensitiesOpenDevice} \textsf{(a)} & $|0\rangle $ & 
$|0\rangle +|1\rangle $ & $|0\rangle +|1\rangle $
& $|1\rangle $ \\ 
 Fig.~\ref{DensitiesOpenDevice} \textsf{(b)} & $|0\rangle $ & 
 $|0\rangle +|1\rangle $ & $(1+i)|0\rangle
+(1-i)|1\rangle $ & $i|0\rangle +|1\rangle $ \\ 
 Fig.~\ref{DensitiesOpenDevice} \textsf{(c)}  & $|0\rangle $ & 
 $|0\rangle +|1\rangle $ & $i|0\rangle -i|1\rangle 
$ & $i|0\rangle $ \\ \hline\hline
\end{tabular}
\caption{Quantum state of the Mach-Zehnder qubit in the four device 
segments indicated in Fig.~\ref{QWRqubitDEF} (columns), corresponding to
the three situations shown in Fig.~\ref{DensitiesOpenDevice} (rows), 
respectively. } 
\label{tableQWRqubit} 
\end{center}
\end{table}

\begin{figure}
\begin{center}
\includegraphics{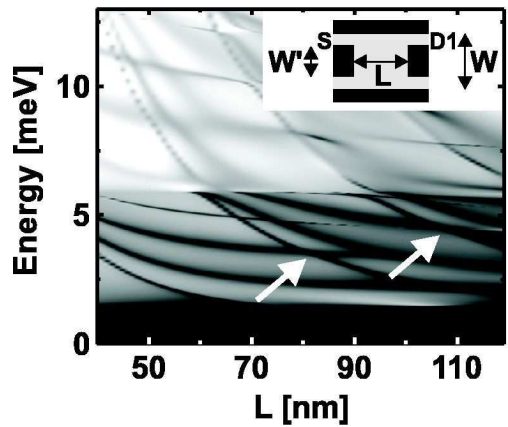}
\caption{Contour plot of the transmission function $T_1$ between source 
$S$ and drain $D_1$ as a function of kinetic energy in meV and as a 
function of the coupling window length $L$ in nm. The total width of the 
structure equals $W=220$~nm, whereas the width of the inner coupling 
windows amounts to 
$W^{\prime}=100$~nm. The darker the color, the smaller the value of $T_1$. The
dark lines therefore indicate minima in $T_1$. The arrows point to 
crossings of resonances where $T_1$ vanishes.}
\label{TransmissionHStructure}
\end{center}
\end{figure}

\begin{figure}
\begin{center}
\includegraphics{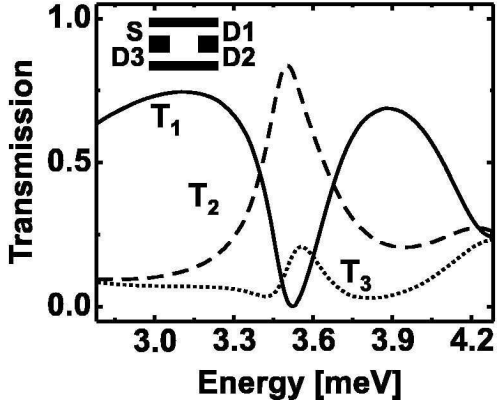}
\caption{Transmission functions 
$T_1$ (solid), $T_2$ (dashed), and $T_3$ (dotted),
as a function of the kinetic energy of the electron in meV. The length $L$ 
of the coupling window equal has been set to $82~\mathrm{nm}$. 
The other symbols have
been defined in Figure~\ref{SchematicDevice3D}.}
\label{TransmissionResonance}
\end{center}
\end{figure}

Since it is unlikely that one can fabricate the device discussed in 
Fig.~\ref{HeterostructureAndMetalGates} precisely, it is imperative to 
understand the dependence of the Mach-Zehnder interferences on the various 
device geometry parameters such as width and length of 
the QWRs and the coupling windows not only quantitatively but also 
qualitatively. A physically transparent picture can be obtained by 
studying a model H shaped structure that 
consists of a single coupling window where the depletion potentials are 
hard-wall potentials as depicted in the inset of 
Fig.~\ref{TransmissionHStructure}.

The coupling between the two QWRs causes the transmission to become 
resonant for specific energies that depend on the length $L$ and 
width $W$ of the coupling window. These energies are approximately equal 
to the energies of the bound states in a rectangular box with the 
dimensions of the coupling window, i.e.\
\begin{equation}
E(n,m)\propto\frac{n^{2}}{L^{2}}+\frac{m^{2}}{W^{2}}.
\end{equation}
Beam splitting, i.e.\ a crossover in the transmission from channel 
$S$-$D_1$ to $S$-$D_2$, requires that the corresponding transmission 
functions obey the relation 
$T_{1}(E_{0})\approx T_{2}(E_{0})\approx0.5$ for some energy $E_{0}$. 
In turn, this condition requires that there are 2 sets of values $(n,m)$ 
that correspond to the same energy $E$, i.e. 
$E(n_{1},m_{1})=E(n_{2},m_{2})$. Thus, the crucial geometric 
requirement for quantum mechanical switching is to achieve an appropriate
ratio of length $L$ and width $W$ that obeys this condition. 
In order to obtain a pronounced resonance, the values of $n,m$ should be 
of the order of unity additionally, since the resonances lie 
densely in energy for 
large integers. Figure~\ref{TransmissionHStructure} shows a contour 
graph of the transmission probability $T_{1}$ (from source $S$ to 
drain $D_{1}$) as a function of the 
coupling window length $L$ and the kinetic energy of the electron. The
arrows in this figure point to suitable values of the length $L$ where
the energies of two eigenstates cross. To further illustrate this 
crossover in the transmissions, Fig.~\ref{TransmissionResonance} 
shows a cross-section through Fig.~\ref{TransmissionHStructure} 
for the length $L=82~\mathrm{nm}$ that is marked by the leftmost
white arrow. This crossing corresponds to the 
switching from channel $S$-$D_1$ 
to $S$-$D_2$ and occurs near the degeneracy of the energy states
$E(1,3)$ and $E(2,1)$.

\subsection{Results: Entangled Mach-Zehnder double 
quantum dot device (2-qubit device)}

In the following, we present the results for the entangled Mach-Zehnder 
double quantum dot device that has been schematically depicted in 
Fig.~\ref{SchematicDevice3D}.  For these calculations, 
we have used a somewhat different set of geometry parameters for the 
Mach-Zehnder interferometer in order to optimize the entanglement. Since 
the calculations of the entangled system are very demanding, we have 
defined the Mach-Zehnder interferometer by a hard-wall potential within 
a $10~\mathrm{nm}$ thick slab of GaAs rather than performing a charge 
self-consistent calculation including all metal gates. We have taken the 
two QWRs to be $55~\mathrm{nm}$ wide, $1000~\mathrm{nm}$ long, and 
the lateral distance between them has been set to $20~\mathrm{nm}$. 
The coupling windows now have a length of $85~\mathrm{nm}$. The resulting 
energy spacing of the two lowest subbands amounts to $5.3~\mathrm{meV}$ 
which is nearly twice as large as in the Mach-Zehnder device of 
Sec.~\ref{results}~A. The Fermi level $E_{\mathrm{F}}$ has been set 
to $1.6~\mathrm{meV}$ in the lowest subband. This Fermi level causes the 
two coupling windows to act as almost perfect beam-splitters, with 
channel transmissions 
$T_{1}(E_{\mathrm{F} })\approx T_{2}(E_{\mathrm{F}})\approx 0.5$. 
For the calculation of the interaction between the Mach-Zehnder 
interferometer and the double quantum dot, we assume a vertical distance 
of $80~\mathrm{nm}$ between them. The larger of the two quantum dots 
with a lower ground state energy is located exactly underneath the center 
of the Mach-Zehnder interferometer whereas the smaller one lies 
$60~\mathrm{nm}$ away in the direction perpendicular to the QWRs. The 
double quantum dot system is modeled by the Hamiltonian given in 
Eq.~(\ref{ModelHamiltonianQuantumDots}). 
The positions of the Mach-Zehnder interferometer and the double quantum 
dot relative to each other have a large influence on the operation 
characteristics of the device. The present configuration exploits the 
idea that the interaction between the electron in the QWRs and the 
electron in the larger quantum dot is identical for both wires. The 
electron in the smaller quantum dot, on the other 
hand, lies closer to the QWR between gates $D_{2},D_{3}$ and interacts 
mostly with the electron in that wire. Consequently, a phase difference 
in the electron wave function for the two QWRs is exclusively induced 
by the electron in the smaller dot.

\begin{figure}
\begin{center}
\includegraphics{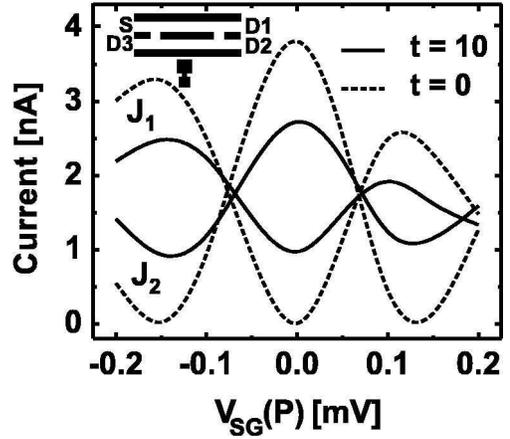}
\caption{Currents $J_{1}$ and $J_{2}$ in nA as a function of the phase 
gate voltage $V_\mathrm{SG}(\mathrm{P})$ in mV for two 
different quantum dot tunneling 
couplings $t$ in $\mathrm{\mu eV}$. The inset is a simplified version 
of Figure~\ref{SchematicDevice3D}. }
\label{CurrentEntangler}
\end{center}
\end{figure}

At first, we study the transfer characteristics of the device for a 
situation where the tunneling between the quantum dots is 
inhibited ($t=0$). The ground 
states of the larger and the smaller quantum dot are assumed to differ by 
$\Delta=10~\mathrm{\mu eV}$. The resulting current in 
the QWRs is shown in Fig.~\ref{CurrentEntangler} by the dashed curves. We 
obtain the typical interference pattern that we have also found in 
Sec.~\ref{results}~A for the all-electric Mach-Zehnder interferometer. 
The asymmetry with respect to the sign of the gate voltage is caused by 
the asymmetric position of the phase 
gate along only one of the two QWRs in combination with the additional 
repulsive potential due to the electron in the double quantum dot.

\begin{figure}
\begin{center}
\includegraphics{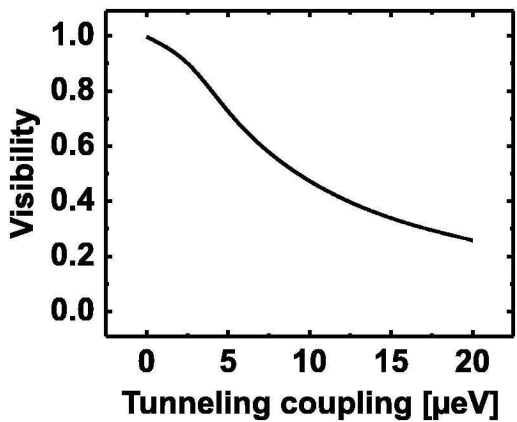}
\caption{Visibility as defined in the text, corresponding to the 
interference pattern of Fig.~\ref{CurrentEntangler}, as a function of the 
tunneling coupling in $\mathrm{\mu eV}$.}
\label{VisibilityVsCoupling}
\end{center}
\end{figure}

\begin{figure}
\begin{center}
\includegraphics{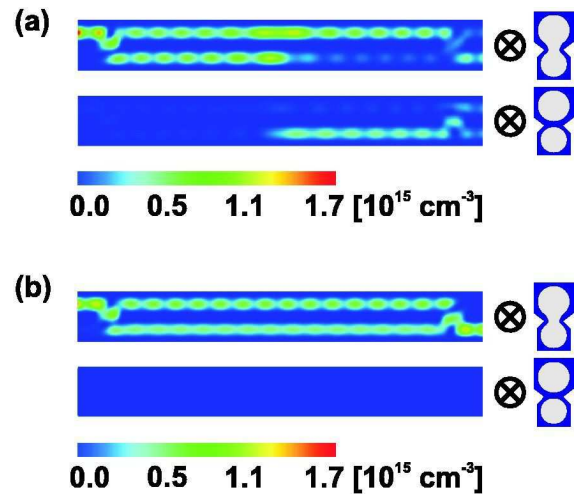}
\caption{(Color online) (a) Contour plot of the stationary charge 
density of the current-carrying states, in units of 
$10^{15}~\mathrm{cm}^{-3}$. The upper (lower) panel shows the 
single-particle electron density in the QWRs when the electron in the 
quantum dot occupies the bonding (anti-bonding) state. The quantum dot 
tunneling coupling 
amounts to $t=10~\mathrm{\mu eV}$. (b) Same type of figure, but with the 
Coulomb interaction between the electron in the wires and the electron 
in the quantum dot switched off.
In case (b), the two possible paths of the scattered electron can
interfere with one another. In case (a), however, the electron
propagating in the lower wire changes the double-dot state and looses
its capability to interfere with the other path, thus reducing the
visibility (see text).
}
\label{EntangledDensity}
\end{center}
\end{figure}

We now allow the electron in the double quantum dot to tunnel between the 
two dots ($t=10~\mathrm{\mu eV}$).  The currents $J_{1}$ and $J_{2}$ that 
correspond to this situation are shown in Fig.~\ref{CurrentEntangler} by 
the solid curves. We still obtain the Mach-Zehnder interference pattern, 
but the visibility is markedly reduced.  Let us define the visibility by
\begin{equation}
v=\left.  \left(  J_{2}-J_{1}\right)  /\left(  J_{2}+J_{1}\right)  \right\vert
_{V_\mathrm{SG}(\mathrm{P})=0}, \label{visibility}
\end{equation}
i.e.\ for zero phase gate voltage $V_\mathrm{SG}(\mathrm{P})$. 
Figure~\ref{VisibilityVsCoupling} shows the visibility as a function of 
the tunneling coupling. The visibility is almost $1$ for vanishing 
tunneling coupling and decreases monotonously with increasing tunneling. 
The visibility is therefore uniquely related to the magnitude of the 
tunneling coupling. The above effect can be explained as follows.  If 
the tunneling probability is high, the double-dot electron can easily 
change its state. In this
regime, it is also easy for the double-dot to \emph{measure} the path
of the interferometer particle, thus causing decoherence and
destroying the interference.  For low tunneling probability, on the other 
hand, the electron that sits initially in the large
dot is insensitive to the Coulomb interaction with the interferometer.
When the tunneling is completely suppressed, the state of the dot is
fixed and the interferometer electron is unable to change the dot state.
While the electrons are still interacting due to their charge in this case, 
the path of the interferometer particle is not measured by the double-dot.

Figures~\ref{EntangledDensity}~(a) and (b) show the electron density 
Eq.~(\ref{ProjectedDensity}) in the entangled Mach-Zehnder double quantum 
dot device. To be precise, the figures depict the projection of the 
probability densities of the current-carrying two-particle states onto 
the eigenstates of the isolated double quantum dot for a tunneling 
coupling of $t=10~\mathrm{\mu eV}$ 
and vanishing phase gate voltage $V_\mathrm{SG}(\mathrm{P})$. In 
particular, Fig.~\ref{EntangledDensity}~(a) shows the interacting case, 
whereas Fig.~\ref{EntangledDensity}~(b) shows the non-interacting case 
where the electron-electron interaction between the electrons in the 
Mach-Zehnder interferometer and the quantum dot electron has been 
switched off. Thus, the two figures illustrate the effect of the Coulomb
interaction on the charge density. In each of the two figures, the upper 
panel corresponds to the ground state, the lower panel corresponds to the 
excited state of the double quantum dot, respectively. In the 
interacting case (a), we find a finite probability of the electron in 
the double quantum dot to be in the excited state. This has to be 
compared to the case of vanishing interaction (b) as well as to the 
case of zero tunneling coupling. In each of the latter cases, 
the probability of the electron in the double quantum dot to be in the 
excited state is zero. This is shown in Fig.~\ref{EntangledDensity}~(b). 
Note that for cases (a) and (b), the total probability densities 
that are given by the sum of the two projected probability densities 
are almost identical near the left contacts, but differ significantly 
near the right contacts. In (a), the total probability density near the 
upper right contact is only partially extinct by 
interference, whereas in (b), the total probability density vanishes 
near the upper right contact. This indicates that the 
interaction between the Mach-Zehnder interferometer and the double 
quantum dot destroys the self-interference of the QWR-electron wave 
function.

\subsection{Discussion: Visibility and von Neumann entropy of the
entangled Mach-Zehnder double quantum dot device}

\begin{figure}
\begin{center} 
\includegraphics{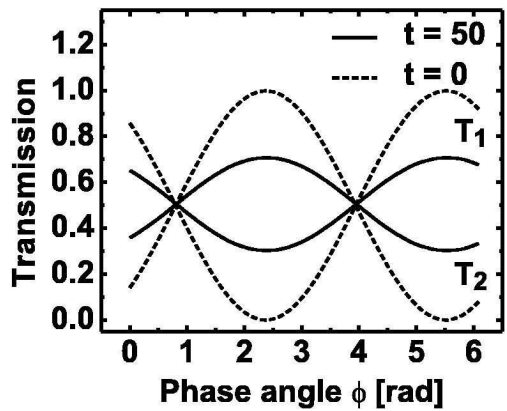} 
\caption{Transmission functions $T_1$ and $T_2$ as a function of 
the phase angle in rad for the analytical model of the entangled 
Mach-Zehnder double quantum dot device. The full line is for a finite 
tunneling coupling $t=50$ in units of $\hbar^2/2mL^2$, whereas the dashed 
curve is for vanishing tunneling coupling that corresponds to no 
entanglement.} 
\label{TransmissionAnalytical} 
\end{center} 
\end{figure}

The results that we have obtained for the entangled Mach-Zehnder
double quantum dot device are indeed the result of a two-qubit quantum
operation.  In order to show this, we apply the analytical model
developed in Sec.~\ref{analyticalmodel} to our two-qubit device.
Figure~\ref{TransmissionAnalytical} shows the transmission
probabilities $T_{1}$ and $T_{2}$, computed from
Eqs.~(\ref{transmission1}) and (\ref{transmission2}), for two
different values of the quantum dots tunneling.  Indeed, the
comparison with Fig.~\ref{CurrentEntangler} shows that the model is
able to reproduce the basic physics of the device and to yield
transmission probabilities $T_{1}$ and $T_{2}$ that qualitatively
agree with the results obtained by the numerical simulations described
in Secs.~\ref{method} and \ref{numericaldetails}.  In
particular, Figs.~\ref{CurrentEntangler} and \ref{TransmissionAnalytical}
show a similar dependence of the visibility on
the tunneling coupling.  The reason for the suppression of the
visibility is the entanglement of the two qubits.  In fact, from the
point of view of the individual qubit, its entanglement with the other
one is nothing else but decoherence and this decoherence partially
suppresses the interference of the electron wave function in the two
QWRs.  In other words, the reduced density matrix of each of the
qubits, obtained by tracing out the degrees of freedom of the other
qubit, represents a mixed state if and only if the two qubits are
entangled.  Thus, the observed suppression of the visibility is both a
direct measure of the degree of entanglement between two qubits and of
the decoherence undergone by the QWRs qubit.

We stress that our approach does not include other sources of
decoherence as, for example, electron-phonon interaction or charge
fluctuations in the metallic split gates that define the structure.
Therefore, our reduction of the visibility is obviously ascribed to the
quantum entanglement of the two qubits.  On the other hand, in real
experiments the visibility will be reduced by the coupling of the
interferometer with any external degree of freedom and not only with
the other qubit.  What makes our results functional to experiments is
the estimation of the visibility decrease against tunable parameters
of the realistic device, as, for example, the tunneling coupling of
the second qubit.  In fact, it is unlikely that the effect of the
environment will be tuned by the same parameters that affect the amount
of entanglement between the two qubits. The phonon effects on the 
interferometer, for example, are determined solely by the temperature,
the structure composition and the energy of the traveling electrons.
Thus, our proposed device demonstrates -- from a modeling
perspective -- a controlled generation of two-qubit entanglement. 
The predicted behavior of the visibility against various system
parameters may serve as an entanglement witness in experimental 
realizations of the system.

In the remainder of this section, we show that the relation between the
visibility of our device, as defined in Eq.~(\ref{visibility}) and
depicted in Figs.~\ref{VisibilityVsCoupling} and
\ref{TransmissionAnalytical}, and the entanglement can be
quantified in terms of the von Neumann entropy of the reduced density
matrix~\cite{PeresBook,NielsenChuang} that is a measure of the
degree of entanglement.

To simplify the derivation, we generally assume that back-scattering
is negligible in the directional couplers and phase gates. This
reduces all matrices to size $4\times4$ and allows the analytical
evaluation of the transmissions $T_1$ and $T_2$ in
Eqs.~(\ref{transmission1}) and (\ref{transmission2}). The density
matrix of the final state is then given by
\begin{equation}
\rho=\sum_{\sigma,\sigma^{\prime}=0,1}\sum_{Y,Y^{\prime}=A,B}E_{\sigma,Y}
^{r}E_{\sigma^{\prime},Y^{\prime}}^{r\ast}\left\vert \sigma\right\rangle
^{\mathrm{W}}\left\vert Y\right\rangle ^{\mathrm{D}}\left\langle
\sigma^{\prime}\right\vert ^{\mathrm{W}}\left\langle Y^{\prime}\right\vert
^{\mathrm{D}}.
\end{equation}
The amount of entanglement of the final two-qubit state is given by
the von Neumann entropy of the reduced density matrix, namely
\begin{equation}
S=-\mathrm{Tr}\,\rho_{\mathrm{red}}\log\rho_{\mathrm{red}}\, ,
\end{equation}
where the reduced density matrix $\rho_{\mathrm{red}}$ is obtained
from $\rho$ by tracing out the electron in the double quantum
dot. This yields the $2\times2$ matrix
\begin{equation}
\rho_{\mathrm{red}}=
\begin{pmatrix}
T_{1} & Z\\
Z^{\ast} & T_{2}
\end{pmatrix}
,
\end{equation}
where the so-called coherence $Z$ is given by
\begin{equation}
Z=E_{0,A}^{r}E_{1,A}^{r\ast}+E_{0,B}^{r}E_{1,B}^{r\ast}.
\end{equation}
With the eigenvalues of the reduced density matrix,
\begin{equation}
p_{\pm}=\frac{1}{2}\left(  1\pm\sqrt{4\left(  \left\vert Z\right\vert
^{2}-T_{1}T_{2}\right)  +1}\right),  \label{EigenvaluesRedDensMatrix}
\end{equation}
the von Neumann entropy reads
\begin{equation}
S=-\sum_{i=+,-}p_{i}\log p_{i}. \label{EntropyDiagonal}
\end{equation}
We now show that the eigenvalues $p_{\pm}$ of $\rho_{\mathrm{red}}$ can be 
expressed in terms of the visibility $v$, Eq.~(\ref{visibility}),
of the interference pattern according to
\begin{equation}
p_{\pm}=\frac{1}{2}\left(  1\pm v\right)  . \label{EigenvaluesVisibility}
\end{equation}
First, we determine the explicit 
expressions for the transmission functions as a function of 
the angle $\phi$ of the phase gate. To this end,
we first apply the $\pi/2$ rotation gate 
$\hat{R}$ and the two-qubit gate $\hat{V}$ onto the initial state 
$\left\vert \Phi_{\mathrm{in}}\right\rangle $. The probabilities  
$|A_{\sigma,Y}|^2$ of the four two-particle basis states 
$\left\vert \sigma\right\rangle ^{\mathrm{W} }\left\vert Y\right\rangle 
^{\mathrm{D}}$ ($\sigma=0,1$, $Y=A,B$) of the resulting state are given by
\begin{equation}
\begin{split}
\left\vert A_{\sigma,B}\right\vert ^{2}  &  =\frac{1}{2}\cos^{2}\left(
\frac{1}{2}\left(  \beta\pm\gamma\right)  \right)  ,\\
\left\vert A_{\sigma,A}\right\vert ^{2}  &  =\frac{1}{2}\sin^{2}\left(
\frac{1}{2}\left(  \beta\pm\gamma\right)  \right) \label{Avisibility} ,
\end{split}
\end{equation}
where $+$ and $-$ correspond to $\sigma=0$ and $\sigma=1$, respectively,
and $\beta$ and $\gamma$ are angular constants that depend on
the two-particle interaction $\hat{V}$. They describe the linear 
combination of the dot-electron states 
$\left\vert B\right\rangle ^{\mathrm{D}}$ and $\left\vert 
A\right\rangle ^{\mathrm{D}}$ within the two-particle wave function. The
explicit expressions of $\beta$ and $\gamma$ do not enter the final result
of this section. The expressions Eq.~(\ref{Avisibility}) reflect the fact 
that the probability for the electron in the QWR to be either in state 
$\left\vert 0\right\rangle^{\mathrm{W}}$ or in state 
$\left\vert 1\right\rangle ^{\mathrm{W}}$ are equal.
The next step consists in applying the phase gate $\hat{P}$ 
and the second $\pi/2$ rotation gate $\hat{R}$. This procedure 
yields the final state probability amplitudes
\begin{equation}
\begin{split}
E_{0,Y}^{r} & =\frac{1}{\sqrt{2}}\left(  \exp\left(  i\phi\right) 
 A_{0,Y} +A_{1,Y}\right)  ,\\
E_{1,Y}^{r} & =\frac{1}{\sqrt{2}}\left(  -\exp\left(  i\phi\right) 
 A_{0,Y} +A_{1,Y}\right)  .
\end{split}
\end{equation}
Note that the gates $\hat{R}$ and $\hat{P}$, being single-qubit 
transformations, 
do not change the degree of entanglement of the two electrons. For the 
transmission probabilities $T_{1}$ and $T_{2}$ we finally obtain
\begin{equation}
\begin{split}
T_{1,2}=\frac{1}{2} & \pm\left\vert A_{0,B}\right\vert \left\vert A_{1,B}
\right\vert \cos\left(  \phi+\delta_{B}\right) \\ 
& \pm\left\vert A_{0,A}
\right\vert \left\vert A_{1,A}\right\vert \cos\left( \phi+\delta_{A}\right)
, \label{TransmissionBasisStates}
\end{split}
\end{equation}
where the signs $\pm$ correspond to $T_{1}$ and $T_{2}$, respectively. 
The phase angles are defined by $\delta_{A} = \arg(A_{0,A}) - \arg(A_{1,A})$ 
and analogously for $\delta_{B}$.
By inserting the expressions for $A_{\sigma,Y}$ into 
Eq.~(\ref{TransmissionBasisStates}), the transmission probabilities read
\begin{equation}
T_{1,2}=\frac{1}{2}\left[  1\pm\sqrt{\cos^{2}\beta\sin^{2}\delta+\cos
^{2}\gamma\cos^{2}\delta}\cos\left(  \phi+\phi_0\right)  \right]  ,
\end{equation}
where $\delta=\frac{1}{2}\left(  \delta_{B}-\delta_{A}\right)$ 
and $\phi_0$ is a constant phase shift. This result shows that the 
transmission probabilities $T_{1}$ and $T_{2}$ yield a sinusoidal 
oscillation as a function of the phase angle $\phi$. By comparing the 
definition Eq.~(\ref{visibility}) of the visibility with
this result for the transmission probabilities, we see that the 
visibility can be written in the form
\begin{equation}
v=\sqrt{\cos^{2}\beta\sin^{2}\delta+\cos^{2}\gamma\cos^{2}\delta}.
\end{equation}
Now, we evaluate the expression $\left\vert Z\right\vert ^{2}-T_{1}T_{2}$ 
within Eq.~(\ref{EigenvaluesRedDensMatrix}), and obtain
\begin{align}
\left\vert Z\right\vert ^{2}-T_{1}T_{2}  &  =\frac{1}{4}\left(  \cos^{2}
\beta\sin^{2}\delta+\cos^{2}\gamma\cos^{2}\delta-1\right) \notag \\
&  =\frac{1}{4}\left(  v^{2}-1\right)  .
\end{align}
Inserting this result into Eq.~(\ref{EigenvaluesRedDensMatrix}), we 
finally obtain the assertion Eq.~(\ref{EigenvaluesVisibility}). 

\begin{figure}
\begin{center}
\includegraphics{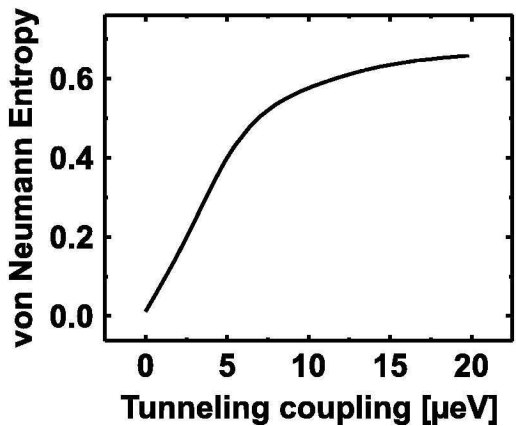}
\caption{Calculated von Neumann entropy, as defined in the text, of the 
entangled Mach-Zehnder double quantum dot device of
Sec.~\ref{results}~C, as a function of the tunneling coupling in
$\mathrm{\mu eV}$.}
\label{EntropyVsCoupling}
\end{center}
\end{figure}

This result proves that the von Neumann entropy of the reduced density
matrix depends monotonously on the visibility.  A visibility of $v=1$ 
corresponds to vanishing von Neumann entropy (zero degree of 
entanglement), while for a visibility of $v=0$, we find a von Neumann 
entropy of $S=1$ (maximum degree of entanglement).  While the relation 
Eq.~(\ref{EigenvaluesVisibility}) between visibility $v$ and von Neumann
entropy $S$ may not hold rigorously for the realistic entangled 
Mach-Zehnder double quantum dot device, it is nevertheless interesting to 
calculate $S$ for this device as a function of the
tunneling coupling, invoking the relation of 
Eq.~(\ref{EigenvaluesVisibility}). The result is presented in 
Fig.~\ref{EntropyVsCoupling} and shows that the
degree of entanglement of the realistic device monotonously increases 
with increasing tunneling coupling.

\section{Conclusion}  \label{conclusions}

We have theoretically analyzed semiconductor single- and 
two-qubit quantum gates based on electrostatically defined QWRs and 
quantum dots. We predict the detailed three-dimensional geometry, 
material composition, doping profile, and bias voltage of an all-electric 
Mach-Zehnder interferometer.  Our calculations of the electronic 
structure and ballistic transport properties of this device show that 
the proposed device is a fully controllable single-qubit gate for 
electrons that propagate in QWRs. The fabrication of the Mach-Zehnder 
interferometer is within the reach of 
present-day technology but full control of the electronic beam-splitters 
requires a very high fabrication precision in the range of a few 
$\mathrm{nm}$.

Based on this all-electric Mach-Zehnder interferometer, we predict a 
two-qubit quantum transport device where the electrons in the 
interferometer couple to a single-electron double quantum dot by Coulomb 
interaction.  We have calculated the ballistic transport properties of 
the three-dimensional two-qubit device with an interacting two-particle 
quantum transport method and designed the device geometry for optimal 
entanglement.  By means of an analytical model of the 
device, we have illustrated the qualitative physics of the two-qubit 
device and showed that the visibility is a faithful measure of the 
entanglement. In particular, we have found that the visibility can be 
controlled externally by tuning the tunneling coupling between the two 
quantum dots of the second qubit. The device realizes a non-trivial 
two-qubit gate that allows the controlled generation and straightforward 
detection of entanglement from DC current-voltage characteristics.

\begin{acknowledgments}
The authors acknowledge support from the Deutsche Forschungsgemeinschaft.
\end{acknowledgments}

\end{document}